\documentclass[twocolumn,prl,amsmath]{revtex4}

\usepackage{graphicx}

\begin{document}

\title
{Accelerated growth in outgoing links in  evolving  networks: deterministic vs. stochastic picture 
}
\author
{
Parongama Sen
}
\affiliation
{
Department of Physics, University of Calcutta,
    92 Acharya Prafulla Chandra Road, Kolkata 700009, India. \\
}
\email {parongama@vsnl.net, paro@cubmb.ernet.in}

\begin{abstract}

In  several real-world networks like the Internet, WWW etc., the number of 
links grow in time in a non-linear fashion. We consider growing networks in 
which the number of outgoing links is a non-linear function of time but new 
links between older nodes are forbidden. The attachments are made using a 
preferential attachment scheme. In the deterministic picture, the number of 
outgoing links $m(t)$ at any time $t$ is taken as $N(t)^\theta$ where $N(t)$ 
is the number of nodes present at that time. The continuum theory predicts a 
power law decay of the degree distribution: $P(k) \propto k^{-1-\frac{2}
{1-\theta}}$, while the  degree of the node introduced at time $t_i$ is given 
by $k(t_i,t) = t_i^{\theta}[ \frac {t}{t_i}]^{\frac {1+\theta}{2}}$ when the 
network is evolved till time $t$. Numerical  results show a growth in the 
degree distribution  for small $k$ values at any non-zero $\theta$. In the 
stochastic picture,  $m(t)$ is a random variable. As long as  $\langle m(t)
\rangle$ is independent of time, the network shows a behaviour similar to the 
Barab\'asi-Albert (BA) model. Different results are obtained when $\langle m(t)
\rangle$ is time-dependent, e.g., when $m(t)$ follows a distribution $P(m) 
\propto m^{-\lambda}$. The behaviour of $P(k)$ changes significantly as $\lambda$ 
is varied: for $\lambda > 3$, the network has a scale-free distribution belonging 
to the BA class as predicted by the mean field  theory, for smaller values of 
$\lambda$ it shows  different behaviour.
Characteristic features of the clustering coefficients in both models have also 
been discussed.

\end{abstract}

\maketitle

\section {I Introduction}

In many real-world networks which evolve in time,  the number of links
show a nonlinear growth in time \cite{doro_review}.  Examples of such network are  the  Internet
\cite{Faloutsos}, World Wide Web (WWW)  \cite{broder},  collaboration \cite{collab}, 
 word web \cite{doro_word},
citation  \cite{redner, citation},  metabolism  \cite{metabolism},
gene regulatory network \cite{gagen1,gagen2} etc.
The number of
links may increase in a two-fold  way: new nodes may tend to get attached to 
more nodes as the
 size of the network increases,
and secondly there may be additional links generated between the older
nodes in a non-linear fashion as shown in Fig. 1. These two factors maybe present  
either singly  or simultaneously resulting in the 
accelerated growth. In some networks 
like the citation and the gene regulatory network, new links  between older nodes
are forbidden and therefore only the  first scheme is valid,  
while in collaboration
network or internet, the second factor is dominating.

The case  when the new node gets a fixed  number of links but
older nodes get new links in a non-linear fashion has been 
considered in both isotropic and directed models of growing networks 
\cite{collab,doro_word,doro1}, 
showing that
it is distinct from networks with a linear growth rule. 
Here, the 
number of links  generated between the older nodes is considered  to have a  power law growth 
in time.
This choice is made because  the assumption  that scale-free behaviour
(which is a desirable feature of networks) is present in a network 
with accelerated growth has been argued  
to be consistent with
a power-law growth of links 
 \cite{doro1}.
 The evolution of the networks in these models of
 accelerated growth was made using  a preferential
attachment scheme as in the Barab\'asi-Albert network (BA) \cite{BA}. 
In the directed network,  this rule is modified by allowing
an additional parameter, the initial attractiveness, in the attachment rule \cite{doro2}.
There is also an alternative   way of achieving a scale-free network 
proposed by
Huberman and Adamic \cite{HA} which has been  used
for modelling the internet network with accelerated growth\cite{kahng}.

\begin{center}
\begin{figure}
\noindent \includegraphics[clip, width=8cm]{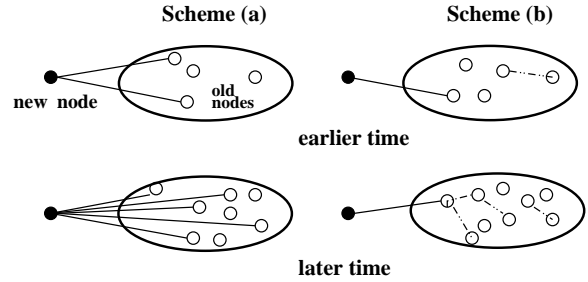}
\caption{ The way accelerated growth takes place: 
in scheme (a), followed in this paper,
the new node gets an increased number of links as time progresses,
no new link between old nodes are allowed. In  scheme (b),
the new node gets a single link and old nodes get new links (shown by dashed lines)
with the total number of links following a nonlinear growth in time.
	The most general case is a combination of  the two schemes.}
\end{figure}
\end{center}
The preferential attachment scheme in the original BA network 
is the simple rule that the incoming nodes get attached to the $i$th node
according to the probability
\begin{equation}
\Pi_i  =  \frac {k_i}{\Sigma k_i},
\end{equation}
where $k_i$ is the degree (number of coonections) of the $i$th node.
This leads to the result that the number of links $k$ is 
distributed according to 
\begin{equation}
P(k) \propto k^{-\gamma}
\end{equation}
for large k.
Let	the $i$th node be introduced at time $t_i$. 
Usually  one node at a time is introduced such that 
$i= t_i$. 
The degree of the $i$th node at a later time $t$ is then denoted by  $ k(t_i,t)$,   
 which in the BA model can be estimated as  
\begin{equation}
        k(t_i,t) = const ~ [\frac{t}{t_i}]^{\beta}.
\end{equation}

In general, the exponent $\beta$ describes  the variation of $k(t_i,t)$
with $t_i^{-1}$, however in the case of the BA model, the behaviour with $t$ 
is also given by the same exponent.
The values of the exponents can be obtained in the BA model  exactly: $\gamma = 3$ and $
\beta = 1/2$, satisfying  the relation $\beta(\gamma-1)=1$. This 
relation holds good in a more generalised version of the
BA model as well \cite{doro2}.
In the BA network, the incoming node gets  a fixed number of links $m_0$  
and the characteristics of the network  do not depend on the specific value of $m_0$.
No new link between older nodes is allowed.
The models of accelerated growth considered so far assume 
that older nodes  get new links (scheme (b) in Fig. 1), which is a sufficient departure from
the original Barab\'asi-Albert model. 
We  consider the simpler case where the number of 
outgoing links  $m(t)$ is no longer a constant but a function of time
and no new link between older nodes is allowed (scheme (a) of Fig. 1).
This is also a realistic scheme as one can expect the number of
attachments of a new node to increase when it is exposed 
to a larger environment.
We have considered both deterministic and stochastic
rules governing the form of $m(t)$. 

The focus of the present paper is on the behaviour of the
various degree distributions in the networks with accelerated growth. 
A brief discussion of the 
clustering properties of the networks  have also been made.
The clustering coefficient ${\cal {C}}{_i}$ measures the 
number of links between the neighbours of the $i$th node.
Here we have estimated the average clustering coefficient 
${\cal {C}}(t)$ in a network evolved upto time $t$ (equal to the
number of nodes) and also the
the average  clustering coefficient  ${\cal {C}}(k)$ 
of nodes with degree $k$. These quantities have been shown to have 
interesting properties in networks \cite{cluster}.

In section II, the deterministic picture is discussed  where 
the number of outgoing links increases  in time 
in a deterministic manner
 and in section III,
stochastic models are considered in which the number of outgoing
links is a random variable. In the last section, the results are
summarised and discussed.

\section {II Deterministic Model}

Let the number of nodes in a growing network be $N(t)$ at time $t$.
In the deterministic model, we take the number of 
outgoing links available to an incoming node to be $m(t) = N(t)^\theta$.
In a  network which is grown by introducing one node at a time,
$N(t) = t$, and therefore $m(t) = t^\theta$, ensuring 
an accelerated growth in the number of links in the network.
The links are made according to equation (1) as usual.

One can obtain an expression for  
$k(t_i,t)$ and $P(k)$   
 using a continuum theory following \cite{continuum}.
Here the rate of change of $k(t_i,t)$  is taken proportional to
$k_i/\Sigma k_i$. Going to the continuum limit 
the total number of links is $\int t^\theta d\theta  = 
t^{1+\theta}/(1+\theta)$
at  time $t$,
and the equation governing $k(t_i,t)$  takes the form
\begin{equation}
\frac {\partial k(t_i,t)}{\partial t} =  \frac {(\theta +1)k(t_i,t)}{2t},
 \end{equation}
leading to
 \begin{equation}
 k(t_i,t) =  t_i^{\theta}[\frac {t}{t_i}] ^{(1+\theta)/2}. 
  \end{equation}
In the last equation the boundary condition $k(t_i, t=t_i) =  t_i^{\theta}$
  has been inserted.
From the above equation, we find that  $\beta = (1-\theta)/2$.
That  $k(t_i,t)$ is not a function of $t/t_i$ alone is a result significantly
different from the BA network.
The degree distribution at time $t$ shows the following behaviour:
\begin{equation}
P(k)  = \frac{2}{1-\theta}  k^{-1 -\frac{2}{1-\theta}} f(t).
\end{equation}
where $f(t) = \frac{1}{1+t} t^\frac {1+\theta}{1-\theta}$.
The value of $\gamma$  is therefore given by 
\begin{equation}
\gamma = 1 + \frac{2}{1-\theta}.
\end{equation}

In this model,  there are two known limits: 
$\theta =0$ corresponds to the BA model and $\theta =1$ corresponds to
a fully connected network (i.e., an $N$-clique). For the BA model, the degree distribution
is a power-law distribution while for the fully connected 
network, $P(k)$ is a delta function. This implies the possibility of the 
existence of a `critical' $\theta_c$ where a peak occurs in the
degree distribution for the first time. 
We conduct numerical simulations to explore 
this possibility.

\begin{center}
\begin{figure}
\noindent \includegraphics[clip,width= 6cm, angle=270]{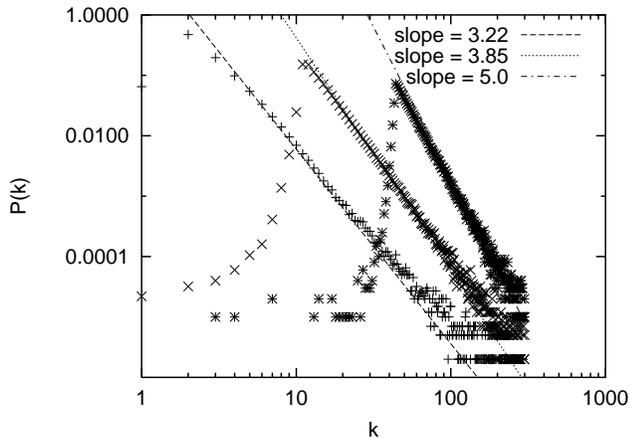}
\caption{ The normalised degree distribution in the deterministic network 
for $\theta = 0.1, 0.3$ and $0.5$ are shown along-with
straight lines in the log-log plot which have  slopes according to 
equation (6). For $\theta = 0.1$ and $0.3$, $t = 4000$
and for $\theta = 0.5$, $t=2000$.}
\end{figure}
\end{center}

In the simulation, nodes are added one by one. A specific number of 
links are assigned to each incoming node ($m(t) = t^\theta$: the nearest 
integer is chosen)
and links are made 
by the preferential attachment scheme.
For larger values of $\theta$ the network becomes highly connected and 
it takes a lot of time to generate it. Hence we restrict to 
values of $\theta \le 0.7$ and to  times $t \le 4000$; the latter
is also the size of the network.
The results show complete agreement with the analytical results
as far as the decay of the degree distribution for large $k$ is concerned
(Fig. 2).
A  growth of the distribution for small $k$ values
is noted as well. This growth seems to be faster than exponential as 
$\theta$ is made larger.
This fast growth suggests that the form of $P(k)$ maybe
assumed to be $P(k) \sim k^{-\gamma} \Theta (k-k_c)$ where $k_c$ is the value at which 
$P(k)$ is maximum. The normalisation  of $P(k)$ 
can then  be done by making  
$\int_{k_c}^\infty P(k) dk = 1$.  Substituting
$P(k)$ from equation (6) and following the above normalisation procedure, 
we get an estimate of $k_c$ as a function of $t$, which is precisely 
$k_c =  t^\theta$ for large $t$.
The numerical results for discrete systems  also agree with the above scaling, the
agreement becoming better
for larger values of $\theta$.
\begin{center}
\begin{figure}
\noindent \includegraphics[clip,width= 6cm, angle=270]{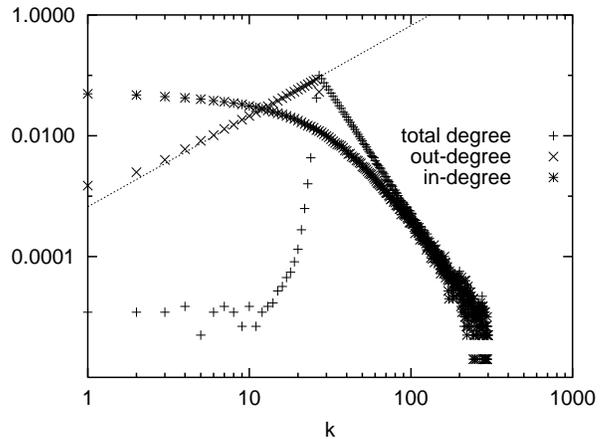}
\caption{The in-degree, out-degree and total degree 
 distributions for $\theta = 0.4$ are shown. $P_{out}(k_{out})$ 
 is fitted with the calculated slope given by eq. 8.}
\end{figure}
\end{center}

Interestingly, we find that a peak is present in the degree distribution 
even for values of $\theta <1$
which  indicates the existence of $\theta_c$.
Analysing the numerical results for small values of $\theta$, we find that
the peak in the distribution occurs as soon as $\theta$ is made non-zero.
This is established by the fact that as the network  size is made larger 
(i.e., the time to which the network is evolved is made larger)  
the  peak  becomes sharper and $P(k)$ decreases  for small $k$ values.
 Hence we conclude that  $\theta_c = 0$.

\begin{center}
\begin{figure}
\noindent \includegraphics[clip,width= 6cm, angle=270]{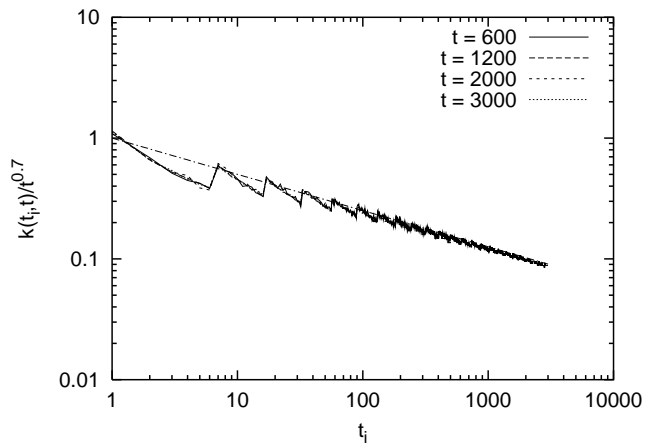}
\caption{
	The scaling plot for $k(t_i,t)$ is shown for $\theta = 0.4$.
	Here $k(t_i,t)$ has been scaled by 
$t^{0.7}$ as $k(t_i,t)$ varies as $t^ {(\theta  +1)/2}$ according to equation (5). 
The straight-line is drawn with slope 0.3 as $\beta = (1-\theta)/2= 0.3$ here.}
\end{figure}
\end{center}

Although in this network $P(k)$ has been calculated 
from the total degree of the nodes, we notice an interesting behaviour of the 
 distributions of the in-degree $k_{in}$ and 
out-degree $k_{out}$ taken separately. 
It maybe mentioned here that these distributions have also
been obtained for real networks and in many of them it is seen
that these are also scale-free with distinct exponents \cite{review}.
The number of outgoing links in the present
network is deterministic and has the following feature: the first $n_1$ nodes
have out-degree 1, the next $n_2$ have out-degree 2 and so on (this is 
due to the discrete nature of the network). For $\theta = 0$,
all nodes have out-degree 1 and its distribution $P_{out}(k_{out})$
is a delta function ($n_1$ equals the number
of nodes in the system and all $n_i = 0$ for $i \ne 1$). For $\theta=1$, 
$n_1 =1,  n_2 = 1, n_3 = 1$ etc., and the 
out-degree distribution is a flat one. For intermediate values of $\theta$,
the exact form of $P_{out}(k_{out}) $ can be easily 
found out. Let $t$ be the first time when the out-degree
of the incoming node is $k_{out}$ and $t + \Delta t$ the time at which
the out-degree increases to  $k_{out} + 1$. Clearly $\Delta t = P_{out}(k_{out})$ and
since $k_{out} = t^\theta$, we have 
$\theta  t^{\theta -1}\Delta t  = 1$.
Therefore
\begin{equation}
P_{out}(k_{out}) \propto \frac{1}{\theta} k_{out}^{(1-\theta)/\theta}.
\end{equation}
Hence we find that the out-degree distribution actually grows with
the degree, a result which is also verified in the numerical simulations (Fig. 3).
Since for $\theta < 1$, the out-degree never assumes a  very large value in a 
network of finite size, we also note that for large $k$, the contribution to $P(k)$
is mainly from $k_{in}$. Consequently, we expect that $P_{in}(k_{in})$
will have a power-law tail with $\gamma$ given by equation (7).
This is also confirmed numerically. The growing region of the total
degree distribution for small $k$ is accounted for by the growing out-degree
distribution as in this region both $k_{in}$ and $k_{out}$ 
contribute.

The behaviour of $k(t_i,t)$ is also in complete agreement  with the 
theoretical results:  
plotting the scaled $k(t_i,t)/t^{(1+\theta)/2}$ against $t_i$ for  
different values of $t$, a  data collapse is obtained. This 
is shown in Fig. 4 for $\theta = 0.4$. 
The agreement with the continuum results becomes  better as $t_i$ increases.

All clustering coeffcients at $\theta=0$  are zero as 
no loops are allowed in this case. 
As $\theta$ is increased, ${\cal {C}}(t)$  (with fixed $t$) shows an increase as 
expected. The increase is not very sharp at small 
values of $\theta$, e.g., for $\theta$ as high as 0.5,  ${\cal {C}}(t)= 0.038$
for $t=2000$. Since ${\cal {C}}(t)=1$ for $\theta = 1$, 
it is expected that ${\cal {C}}(t)$ will show a sharper  increase  
for larger values of $\theta$; it is however difficult to simulate 
networks in this range of  values of $\theta$ to check this behaviour.
${\cal {C}}(t)$ as a function of $t$  shows a behaviour similar to the 
BA model; it decreases with $t$  (at least upto $\theta = 0.4$; 
we have not studied this variation beyond this value). 
This decrease is also expected to have a dependence on 
$\theta$ (in the limit $\theta = 1$, ${\cal {C}}(t)$ 
is independent 
of $t$). We have not, however, attempted to study in detail
the dependence of ${\cal {C}}(t)$ with $t$ 
as  $\theta$ is varied.

We have also calculated ${\cal {C}}(k)$ when $t$ is kept fixed 
which  again shows a behaviour similar to the BA model for large
values of $k$, i.e.,  for non-zero
values of 
$\theta$, ${\cal {C}}(k)$ becomes a constant.
This constant is a function of $\theta$ and we find that 
${\cal {C}}(k) \sim \theta ^2$  for large $k$ gives a good fit to the data.

\section {III Stochastic models}

The assumption that an incoming node gets attached to a fixed 
number of nodes at a given time  in  the deterministic model is
somewhat artificial. In most social networks, 
the outgoing links also have a distribution
which  usually   shows a decay \cite{review}.
Hence one should  consider 
randomness in the number of outgoing links in a realistic manner
such that  the number of outgoing links $m$ is not fixed at time $t$
but is a stochastic variable.
In fact, $m$ can be a stochastic variable even in the 
conventional BA model by   not putting  any restriction
on the number of outgoing links. This can be achieved  by simply 
allowing each existing node  the possibility to get attached to  the
incoming node
according to the probability given in equation (1).
However, it is  known  that making $m$ random
in this way does not change the  BA universality class.
This happens because even though $m$ is random, 
 $ \langle m(t) \rangle$, the mean value is practically independent 
 of time. Thus it is possible to replace  $m_0$ by
  $ \langle m(t) \rangle$ in the rate equation for $k(t_i,t)$ \cite{continuum} and get
  the same results as in the BA model. 
Such a replacement is meaningful as long as the fluctuations are
negligible which is true in the unrestricted BA case. This  we have checked
by numerical simulations also.

To get a stochastic model 
in which $\langle m(t)\rangle$ is a function of time, we let   
  $m(t)$ 
 follow a distribution
which depends on the
number of nodes $N(t)$ present at that time.
The choice of the distribution can be done in many ways. However,
we find that in many real networks, where the distribution
of the out-degree has been done, the distribution
shows either a power law (e.g. the WWW or phone-call network) or an exponential 
(e.g. as in the citation network) tail \cite{review}.
This study has been done in networks evolved for a sufficient
duration  of time, here we assume that the same kind of distribution
is valid for intermediate times.
The dependence on the number of nodes present in the system
at time $t$ occurs by putting the upper bound of $m(t)$ equal to $N(t)$.

Taking an exponentially decaying distribution of $m(t)$, however, again
gives nothing new. This is because the mean value $ \langle m(t)\rangle$
 becomes time independent (within  a short  time) and therefore we get the  
BA network again. Thus we focus our attention on the model
in which
 $m(t)$
 follows a power law distribution, i.e.,
 $P_m(m(t)) \propto m(t)^{-\lambda}$ with $1 \leq m(t) \leq N(t)$.
Again we take $N(t) = t$, i.e., one node is being added at a time.

 In this model $ \langle m (t) \rangle$ will have the following behaviour:

\begin{subequations}
\begin{eqnarray}
\langle m(t) \rangle  &\sim &t  ~~~ ({\lambda < 1})\\
  &\sim &t^{2-\lambda}    ~~~({ 1 < \lambda < 2})\\
    &\sim & O(1)  ~~~ ({  \lambda > 2}).
    \end{eqnarray}
    \end{subequations}

\begin{center}
\begin{figure}
\noindent \includegraphics[clip,width= 6cm, angle=270]{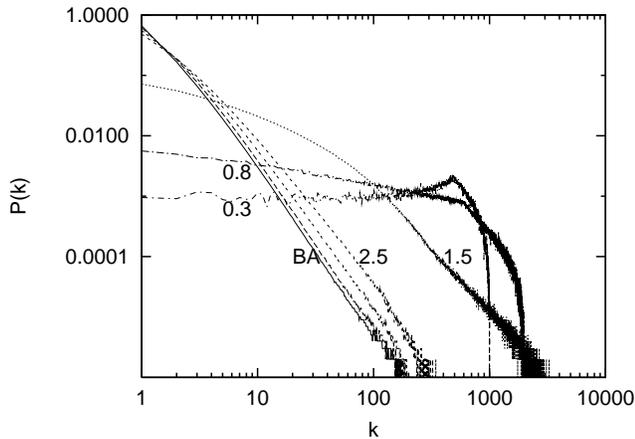}
\caption{The degree  
 distribution for the stochastic model for several values of $\lambda $  (denoted by the
 labels) are shown. In between the BA curve and the curve for $\lambda = 2.5$, the
 two unlabelled curves correspond to $\lambda = 4.0$ and $\lambda = 3.0$. For higher values
 of $\lambda$, the network is grown to 4000 nodes while for the lower-most $\lambda$, we have maximum 1000
 nodes. } 
\end{figure}
\end{center}

Assuming that $m$ can be replaced by its mean 
$\langle m(t) \rangle$, the continuum theory discussed
in the last section can be used for the stochastic model as well
once we define an effective
$\theta$
from the above equations. 
Thus $\theta_{eff} = 0$ for
 $\lambda > 2$,     
       $\theta_{eff} = 1$ for
        $\lambda < 1$, and for $1 < \lambda < 2$, $\theta_{eff} = 2-\lambda$. 
We should not, however, expect this `continuum mean field theory' to be valid if the fluctuations
are not negligible. An  estimate of the fluctuation
in $m$ can also be made which shows that it increases as $\lambda $ decreases and 
cannot be neglected for $\lambda <3$. 

\begin{center}
\begin{figure}
\noindent \includegraphics[clip,width= 6cm, angle=270]{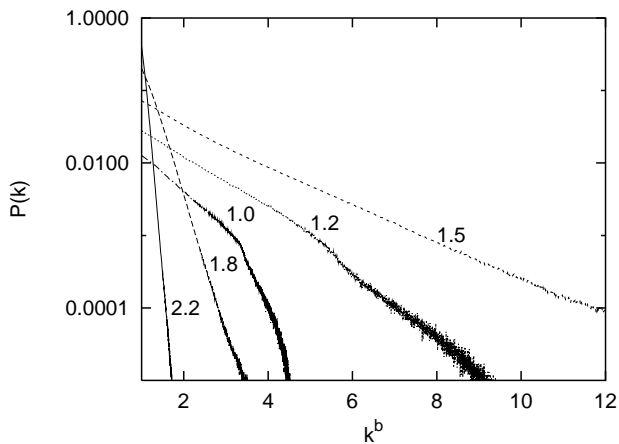}
\caption{The degree  distribution is plotted against $k^b$ to
show the stretched exponential behaviour for several values of $\lambda $ (the labels
denote the values of $\lambda $). The values of $b$ are  0.1, 0.20, 0.42, 0.3 and 0.2
for $\lambda = 2.2, ~ 1.8, ~1.5, ~1.2, ~1.0$  respectively.
} 
\end{figure}
\end{center}

We use numerical simulations to find out the validity of
the `continuum mean field theory' in the stochastic models.
Plotting $P(k)$ vs. $k$ (Fig. 5) for several values of $\lambda$, we find that
for large values of $\lambda $ it is indeed BA-like. As $\lambda$ is decreased
it deviates from the power law behaviour. It is difficult to
locate the exact value of $\lambda $ where the  change in  behaviour occurs,
but clearly, the scale-free behaviour observed for $\lambda > 3$ is no longer
valid for values of $\lambda$  above 2 but less than 3. 
For $2 <\lambda <3$, the fluctuations ($\langle m^2\rangle - \langle m\rangle ^2$) 
scale like $N(t)^{3-\lambda}$, which means
that it becomes stronger for large $N(t)$ and for $\lambda$ closer to
2. Increasing $N(t)$ to large values is difficult
as it takes a long time to generate the network.
So we analyse the behaviour of $P(k)$ for $\lambda$ close to 2, e.g. at $\lambda  = 2.2 $ and $2.5$,
to check the role of fluctuations. For both these values, the
behaviour of $P(k)$ agrees better with a stretched exponential behaviour: $P(k)
\sim \exp(-ak^b)$ albeit with very small values of $b$ ($b \sim 0.10$).
For $1<\lambda <2$,  the deviation from power law is quite clear.
As $\lambda$ is made smaller than 2, the behaviour of $P(k)$ is no longer monotonic
and for small values of $k$ it can be fitted to a stretched exponential function
with $b$ depending on $\lambda$. 
The range of $k$ over  which $P(k)$ follows a stretched exponential behaviour shrinks 
beyond $\lambda = 1.5$ where it shows a sharp drop (see for instance
the curves for $\lambda = 1.2$ and $1.0$ in Fig. 6).

For values of $\lambda <1$, the fluctuations increase rapidly and the
network takes very long time to be generated as it becomes more and more clustered.
$P(k)$ shows a slow  decay  for $0.5 <\lambda < 1$ over a long
range of $k$
and    drops sharply  as $k$ approaches the total
number of nodes in the network. The distribution becomes flatter
as $\lambda $ is decreased.
Below $\lambda = 0.5$, a weakly growing region emerges. 
In fact, a peak appears and becomes sharper as $\lambda$ is made 
smaller. The position of the peak, as $t$ is made larger,
also shifts towards larger $k$, e.g., for $\lambda =0.3$,
the peak is at $k \sim 500$ for $t = 1000$ and at $k \sim 1000$
for $t = 2000$. A systematic study demands a very large amount of 
computer time and we have not attempted it. Nor do we 
try to fit the data to any familiar form because of the 
large amount of fluctuations.

As in case of the deterministic network, here too one can find out
the individual distributions $P_{out}$ and $P_{in}$. As expected,
$P_{out}$ shows a power law decay with exponent $\lambda$. 
$P_{in} $ almost coincides 
with the total degree distribution as the network becomes larger.

The behaviour of $k(t_i,t)$ has also been studied in this network. Again we
find a deviation from the power law behaviour as $\lambda $ becomes less than 
3. As $\lambda$ is made smaller,  $k(t_i,t)$ becomes flat as expected.

The clustering coefficients have been estimated in this model as well.
 ${\cal {C}}(t)$ as a function of $\lambda$ shows a increase 
 as $\lambda$ is decreased as expected (e.g, for $\lambda = 3.0$, 
${\cal {C}}(t) = 0.029$ and for $\lambda = ~1.5$, ${\cal {C}}(t)~ =~  0.679$ for
$t=1000$). 
As a function of $t$, ${\cal {C}}(t)$ shows a decrease with $t$ for $\lambda > 2.0$.
However, as $\lambda$ is made smaller, for the values of $t \leq 2000$
at least, ${\cal {C}}(t)$  shows an 
increase with $t$. We conjecture that ${\cal {C}}(t)$ becomes independent of $t$
for large $t$ values  for  $\lambda < 2.0$.
Here ${\cal {C}}(k)$ shows a logarithmic deacy with $k$: ${\cal {C}}(k) \sim
a - b \log (c k)$ with the values of $a,b$  depending strongly on  $\lambda$
and $c \simeq 1$ for all $\lambda$.
Both $a$ and $b$ approach   zero  for very  large $\lambda$.
As $\lambda$ is decreased both $a$ and $b$ increase indicating
a larger value of ${\cal {C}}(k)$ together with  a sharper decay.

\section{IV Summary, Comments and Discussions}

We have considered both deterministic and stochastic models of growing networks
with preferential attachment in which the number of outgoing
links is a function of time. In both models we have followed simple
rules of evolution with a single tunable parameter. The main results  obtained in the deterministic
model are:\\
1. A power law decay
of the total degree distribution $P(k)$ for large $k$ is obtained 
which is consistent with the 
continuum theory. \\
2. A growing region in $P(k)$ for small $k$ is also observed.
This is due to the behaviour of the out-degree distribution
which has a power law increase. A peak is obtained in
$P(k)$ for any $\theta > 0$.
The peak position $k_c$ is found to vary as  $t^\theta$. \\
3. The values of the exponents $\gamma$ and $\beta$ can be obtained. They
satisfy the relation $\beta(\gamma-1)=1$ as in a general BA model.\\
4. $k(t_i,t)$ can be estimated. It is not a function of $t/t_i$ as in the BA model.\\
5. The degree distribution $P(k)$ is also found to be 
non-stationary, i.e., dependent on the time upto which  the network has been
evolved. \\
The main results of the stochastic model are:\\
1. The behaviour of $P(k)$ depends on the value of $\lambda$. The mean 
field theory predicts a transition at $\lambda =2$ above 
which it becomes BA-like. The numerical
simulations  show that 
for  $\lambda >3$ the degree distribution is scale-free with $\gamma = 3$.\\
2. The degree distribution loses its power law decay nature
at values of $\lambda<3$ and assumes a stretched exponential
form: $P(k) \propto \exp(-ax^b)$.
However, the deviation from the scale-free behaviour is marginal, as indicated by the low
value of $b$ in the region $2 < \lambda  < 3$. Hence this maybe a correction to the BA scaling 
and therefore it may not be correct
to conclude that a phase transition has occurred at $\lambda=3$.
Clearcut inconsistency with the mean field theory is observed for
smaller values of $\lambda$ when the stretched exponential behaviour
of $P(k)$ becomes more pronounced.
For very small values of $\lambda$, a weakly growing region in $P(k)$ emerges.
The data, with a lot of fluctuations, is difficult to handle in this region. However, this
region is not of much interest to us as 
for real networks, such small $\lambda$ values 
seem unrealistic.\\
3. The power law decay behaviour of $k(t_i,t)$ 
observed for $\lambda > 3$ also becomes invalid as $\lambda $ is made smaller.\\
4. For large values of $\lambda$, the degree distribution
is stationary while for small $\lambda$ (presumably for $\lambda < 0.5$) 
it becomes dependent on the size of 
the network.\\

A straightforward comparison of the above two models
shows that the stochastic model is closer to real world   networks. 
This is concluded from comparison
with real world data: the out-degree distributions in phone call,
citation etc. networks show  a decay, either
power law or exponential, while in the deterministic model,
we find a growing behaviour.
Even though we get a power law decay of $P(k)$ in the deterministic
model, the exponent $\gamma$ is always $>3$ while in most 
real networks, it is closer to 2 and can be even less than 2.
The stochastic model on the other hand has a out-degree distribution
which shows a power law decay. It also 
shows that the scale-free behaviour can vanish even in a 
growing network with preferential attachment.
In fact the stretched exponential behaviour of the degree 
distribution observed for small degrees is reminiscent
of the behaviour of the degree distribution in citation network \cite{redner}.
The stochastic model is also of theoretical interest as
it indicates    the  existence of transitions 
at finite values of $\lambda$.
However, the deterministic model  has its own merits: it is
easy to construct and a number of results for it can be obtained
analytically in the continuum limit. The peak obtained in the total degree
distribution in this model  may be compared to
similar peaks obtained in some real world networks e.g., the co-authorship 
network \cite{review} or the Indian railways network \cite{train}.

The citation network is perhaps the most appropriate 
network which the stochastic model emulates. However, 
the citation network is essentially a directed network.
Therefore  we have also considered a specific directed model
in the stochastic case, which follows the attachment rule 
as in \cite{doro2} and where $P_m(m)$ has a peak and an exponential decay
\cite{citation}.
The results show that it has a power law decay in the in-degree
distribution with $\gamma \sim 2$ (not much different
from the case when there is no accelerated growth \cite{doro2}). There are some available data on citation 
network \cite{citation,redner} but the data and the analyses are not
sufficiently general to compare with simulation data.
In this context it may be mentioned that in a model to
simulate the   gene regulatory network, which is  a directed network 
with  accelerated growth, the outgoing links also show a power law 
with exponent close to 2 in good agreement with theory \cite{gagen2}.

In conclusion, we have tried to construct   models of accelerated growth
where the number of outgoing links increase with the network
size  following certain prescriptions. 
A continuum theory can be formulated  for the deterministic case which can 
be applied  in  the stochastic case in the mean field limit.
Based on the different results obtained in the models
we conclude that the stochastic model is closer to real-world
networks.

Acknowledgements: 
The author wishes to thank the networkers of Kolkata: A. Chatterjee,
S. Dasgupta, S. S. Manna, G. Mukherjee and P. Sreeram  for 
comments and encouragement and 
I. Bose and S. Banerji 
for relevant discussions.
Financial support form DST grant SP/S2/M-11/99 is acknowledged.


\end{document}